\documentclass[nofootinbib,showpacs, preprintnumbers, aps,  amsmath, amssymb, preprint]{revtex4-1}

\usepackage{braket}
\usepackage{color}

\usepackage{graphicx}
\usepackage{dcolumn}
\usepackage{bm}
\usepackage[hidelinks]{hyperref}
\usepackage{mathdots}
\usepackage{braket}
\usepackage{MnSymbol}
\usepackage{mathtools}

\newcommand{\ii}{\ensuremath{ \mathrm{i\,} }}
\newcommand{\ee}{\ensuremath{ \mathrm{e}}}

\newcommand{\lu}[1]{{\color{black}{#1}}}

\begin{document}


\title{ Improving the dynamics of quantum sensors with reinforcement learning}

\author{Jonas Schuff}
\author{Lukas J.~Fiderer}%
 \author{Daniel Braun}
\affiliation{%
 Institute for Theoretical Physics, University of T\"ubingen, Auf der Morgenstelle 14, 72076 T\"ubingen, Germany
}%


\date{\today}

\begin{abstract}
Recently proposed \textit{quantum-chaotic sensors} achieve quantum enhancements in measurement precision by applying nonlinear control pulses to the dynamics of the quantum sensor while using classical initial states that are easy to prepare.  Here, we use the cross-entropy method of reinforcement learning to optimize the strength and position of control pulses. Compared to the quantum-chaotic sensors \lu{with periodic control pulses} in the presence of superradiant damping, we find that decoherence can be fought even better and measurement precision can be enhanced further by optimizing the control. In some examples, we find enhancements in sensitivity by more than an order of magnitude. By visualizing the evolution of the quantum state, the mechanism exploited by the reinforcement learning method is identified as a kind of spin-squeezing strategy that is adapted to the superradiant damping. 
\end{abstract}

\maketitle
\onecolumngrid

\tableofcontents
\newpage
\section{Introduction}\label{sec:intro}
The rise of machine learning \cite{murphy2012machine} has led to intense interest in using machine learning in physics, and in particular in combining it with quantum information technology
\cite{dunjko2018machine, mehta2019high}.
Recent success stories include discriminating phases of matter \cite{carrasquilla2017machine,broecker2017quantum,van2017learning} and efficient representation of many-body quantum states \cite{carleo2017solving, carleo2018constructing, gao2017efficient}.

In physics, many problems can be described within control 
theory which is concerned with finding a way to steer a system to achieve a goal \cite{leigh2004control}.
The search for optimal control can naturally be formulated as reinforcement learning \cite{kaelbling1996reinforcement, sutton2018reinforcement, sutton1992reinforcement, chen2013fidelity, palittapongarnpim2017learning, fosel2018reinforcement, bukov2018reinforcement, albarran2018measurement, niu2019universal}, a discipline of machine learning. Reinforcement learning (RL) has been used in the context of quantum control \cite{bukov2018reinforcement}, to design experiments in quantum optics \cite{melnikov2018active}, and
to automatically generate 
sequences of gates and measurements for quantum error correction
\cite{fosel2018reinforcement,sweke2018reinforcement,andreasson2018quantum}.

RL has also been applied to control problems in quantum metrology \cite{dunjko2018machine}: In the context of global parameter estimation, i.e., when the parameter is a priori unknown, the problem of optimizing single-photon adaptive phase-estimation was investigated \cite{hentschel2010machine,hentschel2011efficient,lovett2013differential}.
There, the goal is to estimate an unknown phase difference between the two arms of a Mach--Zehnder interferometer. After each measurement,
 an additional controllable phase in the interferometer can be adjusted dependent on the already acquired measurement outcomes.
 The optimization with respect to policies,  i.e., mappings from measurement outcomes to controlled phase shifts, can be formulated as a RL problem and tackled with particle swarm \cite{hentschel2010machine, hentschel2011efficient,sergeevich2012optimizing,stenberg2016characterization} or differential evolution \cite{lovett2013differential,palittapongarnpim2016controlling} algorithms, where the results of the former were recently applied in an  experiment \cite{lumino2018experimental}.

Also in the regime of local parameter estimation, where the parameter is already known to high precision (typically from previous measurements), actor-critic and proximal-policy-optimization RL algorithms were used to find policies to control the dynamics of quantum sensors \cite{liu2017quantum, liu2017control, xu2019transferable}. There, the estimation of the precession frequency of a dissipative spin-$\frac{1}{2}$ particle was improved by adding a linear control to the dynamics in form of an additional controlled magnetic field \cite{xu2019transferable}.

 Recently it was shown theoretically that the
sensitivity (in the regime of local parameter estimation) of existing quantum sensors based on precession dynamics, such as spin-precession magnetometers, can be increased
by adding nonlinear control to their dynamics in such a way that the dynamics becomes non-regular or (quantum-)chaotic \cite{fiderer2018quantum,fiderer2019quantum}. 
The nonlinear kicks (described by a ``nonlinear'' Hamiltonian $\propto J_y^2$ compared to the ``linear'' precession Hamiltonian $\propto J_z$ where $J_x$, $J_y$, $J_z$ are the spin angular momentum operators)  lead to a torsion, a precession with
rotation angle depending on the state of the spins.

Adding nonlinear kicks to the otherwise regular dynamics comes along with a large number of new degrees of freedom that remained so
far unexplored:  Rather than kicking the system periodically with
always the same strength and with the same preferred axis as in Ref.~\cite{fiderer2018quantum}, one can try to optimize each kick individually,
i.e., vary its timing,  strength, or rotation axis.  The number
of parameters increases linearly with the total measurement time
(assuming a fixed upper bound of kicks per unit time), and becomes
rapidly too large for brute-force optimization.

In this work, we use cross-entropy RL to optimize the kicking strengths and times in order to maximize the quantum Fisher information, whose inverse constitutes a lower bound on the measurement precision. The cross-entropy method is used to train a neural network that takes the current state as input and gives an action on the current state (the nonlinear kicks) as output. In this way, the neural network generates a sequence of kicks that represents the policy for steering the dynamics.

This represents an offline, model-free approach which is aimed at long-term performance, i.e., the optimization is done based on numerical simulations, without being restricted to a specific class of policies, and with the goal of maximizing the quantum Fisher information only after a given time and not, as it would be the case for greedy algorithms, for each time step.
We show that this can lead to largely enhanced sensitivity even compared to the already enhanced sensitivity of the quantum-chaotic sensor with constant periodic kicks \cite{fiderer2018quantum}.
\section{Quantum metrology}
The standard tool for evaluating the sensitivity with which a
parameter can be measured is the quantum Cram\'er-Rao bound
\cite{helstrom1976quantum, Holevo1982, Braunstein94}.  It gives the smallest
uncertainty with which a parameter $\omega$ encoded in a quantum state
(density matrix) $\rho_\omega$ can be estimated.  The bound is
optimized over all possible (POVM=positive operator valued measure)
measurements (including but not limited to standard projective
von-Neumann measurements of quantum observables), and all possible
data-analysis schemes in the sense of using arbitrary unbiased
estimator functions $\hat{\omega}$ of the obtained measurement results. It can be 
saturated in the limit of a large number $M$ of measurements, and
hence gives the ultimate sensitivity that can be reached once technical noise has been eliminated and only the intrinsic
fluctuations due to the quantum state itself remain.    

\begin{figure}[htbp!]
	\centering
	\includegraphics[width=0.9\linewidth]{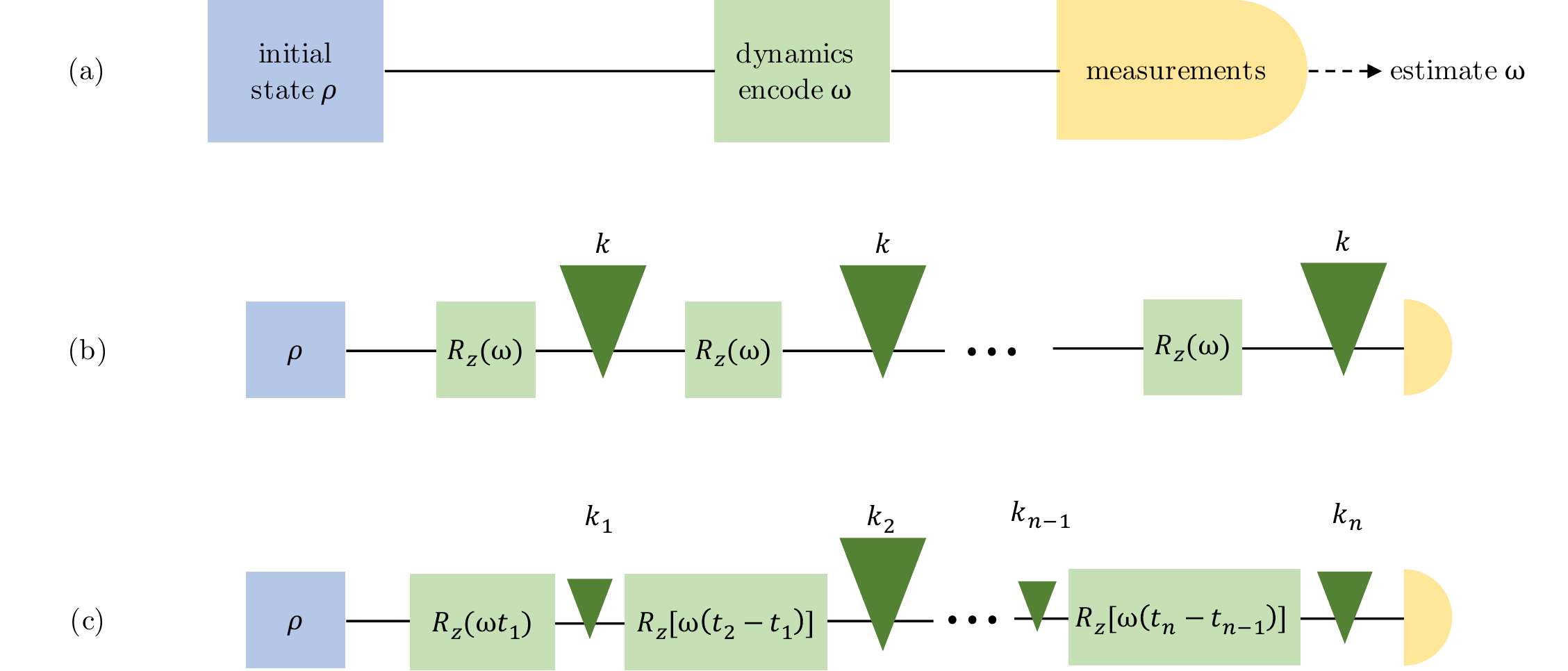}
	\caption{Schematic representation of parameter encoding in quantum metrology. Panel (a) shows the standard protocol: the parameter $\omega$ is encoded in the initial state  $\rho$ through the dynamics, the resulting state is measured, and the parameter is inferred by (classical) post processing of the measurement outcomes. In panel (b), the dynamics is given by the kicked top model:
		the encoding of the parameter $\omega$ through linear precession $R_z(\omega)$ about the $z$-axis is periodically disrupted through parameter-independent, nonlinear, controlled kicks (green triangles) with kicking strength $k$ that can render the dynamics chaotic. In panel (c), the dynamics is given by a generalized kicked top model: the kicking strengths $k_\ell$ and times $t_\ell$ between kicks are optimized in order to maximize the sensitivity with which $\omega$ can be inferred (varying $k_\ell$ are indicated by different sizes of the green triangles). Variation of the kicking axis is possible but beyond the scope of this work.
		}
	\label{fig:qmetro-scheme}
\end{figure}
The quantum Cram\'er-Rao bound for the smallest possible variance of
the estimate $\hat\omega$ reads
\begin{equation}\label{eq:cramer_rao}
\text{Var}(\hat\omega) \geq \frac{1}{M I_\omega}.
\end{equation}
For a state given in diagonalized form,
$\rho_\omega \coloneqq \sum_{\ell=1}^d p_\ell \ket{\psi_\ell}\bra{\psi_\ell}$, where $d$ is the dimension of the Hilbert space, the quantum Fisher information (QFI) is given by \cite{Paris09}
\begin{align}
I_\omega = 2\sum_{\ell, m=1}^{d}\frac{\left|\bra{\psi_\ell}\partial_\omega \rho_\omega\ket{\psi_m}\right|^2}{(p_\ell +p_m)^2},
\end{align}
where the sum runs over all $\ell, m$ such that $p_\ell +p_m\neq 0$, and $\partial_\omega \rho_\omega\coloneqq\frac{\partial\rho_\omega}{\partial\omega}$.
\section{The system}\label{the_system}
We consider a spin model based on the angular momentum algebra, with spin operators $\textbf{J}=(J_x,J_y,J_z)$, $J_z\ket{jm}=\hbar j\ket{jm}$ and $\textbf{J}^2\ket{j,m}=\hbar^2 j(j+1)\ket{j,m}$, where $j$ and $m$ are angular momentum quantum numbers. Note that the model can be implemented not only with physical spins but with any physical system with quantum mechanical operators that fulfill the angular momentum algebra. The Hamiltonian of our model is given by
\begin{equation}\label{eq:hamiltonian}
\mathcal{H}_\text{KT}(t) = \omega J_z + \frac{J_y^2}{(2j+1) \hbar}  \sum_{\ell=-\infty}^{\infty} \kappa_\ell\tau \delta(t-t_\ell)\,.
\end{equation}
The first summand describes a precession about the $z$-axis with precession frequency $\omega$.
The second summand describes the nonlinear kicks, i.e., a torsion about the $y$-axis, see Fig.~\ref{fig:qmetro-scheme}.
This corresponds to a precession about the $y$-axis with a precession angle proportional to the $y$-component.
The time $\tau$ defines a time scale such that  $t$ and $t_\ell$ measure time in units of $\tau$.
The $\ell$th kick is applied at time $t_\ell$ where $\kappa_\ell$ quantifies its kicking strength (in units of a frequency).

In an atomic spin-precession magnetometer, as discussed in Ref.~\cite{fiderer2018quantum}, the first summand corresponds to a Larmor precession characterized by the Larmor frequency $\omega=g\mu_\text{B}B/\hbar$ with Land{\'e} g-factor $g$, Bohr magneton $\mu_\text{B}$, and magnetic field strength $B$, which is the parameter that one wants to estimate. The nonlinear kicks can, for example, be generated with off-resonant light pulses exploiting the ac Stark effect. We introduce a dimensionless kicking strength as $k_\ell\coloneqq \kappa_\ell \tau$ and, for the sake of simplicity, we set $\tau=1$ and $\hbar=1$.

For a pure state, 
the unitary time evolution of the system between kicks at time $t_{\ell-1}$
and $t_\ell$ is given by
\begin{equation}
\ket{\psi_\omega(t_\ell)} = U_\omega( k_\ell)\ket{ \psi (t_{\ell-1})},
\end{equation}
where the unitary transformation $U_\omega( k_\ell)$ propagates the state according to the Hamiltonian \eqref{eq:hamiltonian}, from time $t_{\ell-1}$ [directly after the $(\ell-1)$th kick] to $t_\ell$ [directly after the $\ell$th kick], as indicated by the index $\ell$ [in order to simplify notation, the index $\ell$ of $k$ not only labels the kicking strength at time $t_\ell$ but also refers to the propagation from $t_{\ell-1}$ to $t_\ell$ of $U_\omega( k_\ell)$]. We have
\begin{equation}
U_\omega( k_\ell)= \mathcal{T} \exp{\left[- \ii       \int_{t_{\ell-1}}^{t_\ell}    dt' \mathcal{H}_\text{KT}(t')   \right]},
\end{equation}
where $\mathcal{T}$ denotes time-ordering. Since the kicks are assumed to be instantaneous, this leads to
\begin{align}\label{eq:unitary}
U_\omega(k_\ell)= \exp{\left[-\ii
	k_\ell \frac{J_y^2}{(2j+1)}\right]}  \exp{\left[-\ii \omega (t_\ell-t_{\ell-1})J_z\right]} \,,
\end{align}
i.e., a precession for time $t_\ell-t_{\ell-1}$ followed by a kick of strength $k_\ell$. The kick occurs at the end of the time interval $[t_{\ell-1},t_\ell]$.

For the standard kicked top (KT), see Fig.~\ref{fig:qmetro-scheme}, the kicking strengths are constant, $k_\ell=k$, and kicking times are given by $t_\ell=\ell\tau=\ell$, with $\ell\in\mathbb{N}$. 
Dynamics of the standard KT is non-integrable for $k> 0$ and has a well defined classical limit that shows a transition from regular to chaotic dynamics when $k$ is increased. In Ref.~\cite{fiderer2018quantum} the behavior of the QFI for regular and chaotic dynamics was studied in this transition regime (for $k=3$ and $\omega=\pi/2$) which manifests itself by a mixed classical phase space between regular and chaotic dynamics. Quantum chaos is defined as quantum dynamics that becomes chaotic in the classical limit. In contrast to classical chaos, quantum chaos does not exhibit exponential sensitivity to changes of initial conditions due to the properties of unitary quantum evolution, but can be very sensitive to parameters of the evolution \cite{peres2006quantum}. The kicked top has been realized with atomic spins in a cold gas \cite{chaudhury2009quantum} and with a pair of spin-$\frac{1}{2}$ nuclei using NMR techniques \cite{krithika2019nmr}.
Here, we generalize the standard KT to
kicks of strength $k_\ell$ at arbitrary times $t_\ell$ as given in Eq.~\eqref{eq:unitary}, see also Fig.~\ref{fig:qmetro-scheme}.

Any new quantum metrology method needs to demonstrate its viability in the presence of noise and decoherence. We study two different versions of the KT which differ in the 
decoherence model used: phase damping and superradiant damping.  Both can be
described by Markovian master equations and are well studied models for open quantum systems
\cite{Dicke54,Gross76,Gross82,Braun01B}. While phase damping conserves
the energy and only leads to decoherence in the $\ket{j,m}$ basis,
superradiant damping leads in addition to a relaxation to the ground state $\ket{j,-j}$.
Its combination with periodic kicking in the chaotic regimes is known
to give rise to a non-equilibrium steady state in the form of a
smeared-out strange attractor \cite{Braun01B} that still conserves
information about the parameter $\omega$, whereas without the kicking
the system in presence of superradiant damping simply decays to the
ground state.  The master equations for both processes have the
Kossakowski--Lindblad form \cite{Kossakowski72,Lindblad76}, with  
\begin{align}
\dot{\rho}(t)=\gamma_\text{pd} ([J_z, \rho(t) J_z]+\text{h.c.})\label{eq:pd}  
\end{align}
for phase damping, where
$\dot{\rho}(t)=\frac{\text{d}}{\text{d}t}\rho(t)$, and  
\begin{equation}
\dot{\rho}(t)=\gamma_\text{sr} ([J_-,\rho(t)J_+]+\text{h.c.})\label{eq:sr}
\end{equation}
for {superradiant damping}, where $J_{\pm}\coloneqq J_x\pm i J_y$ are
the ladder operators, and $\gamma_\text{pd}$ and $\gamma_\text{sr}$
denote the  decoherence rates. With the generator $\Lambda$, defined by 
$\dot{\rho}(t)= \Lambda \rho(t)$, one has in both cases the formal
solution $\rho (t_n) = D\left(t_n-t_{n-1}\right)\rho(t_{n-1})$
with the continuous-time propagator $D\left(t\right)\coloneqq \ee^{\Lambda t }$.
The solution of Eq.~\eqref{eq:pd} in the $\ket{j,m}$ basis, where
$\rho(t)=\sum_{m,m'=-j}^j \rho_{m,m'}(t) \ket{j,m}\bra{j,m'}$,  is
immediate,
\begin{align}
\rho_{m,m'}(t)  = \rho_{m,m'}(0) \exp{\left[-\gamma_\text{pd}t(m-m')^2\right]}\,.
\end{align}
Also for Eq.~\eqref{eq:sr} a formally exact solution has been found \cite{bonifacio1971quantum} and  efficient semiclassical (for large $j$)
expressions are available \cite{PBraun98b,PBraun98a}. For our purposes it was the simplest to solve Eq.~\eqref{eq:sr} numerically by
diagonalization of $\Lambda$.
Combining these decoherence mechanisms with the unitary evolution 
the transformation $\rho(t_{\ell-1})\rightarrow\rho(t_\ell)$ reads 
\begin{equation}
\rho(t_\ell)
=U_\omega(k_\ell)\left[D\left(t_\ell-t_{\ell-1}\right)\rho(t_{\ell-1})\right]{U_\omega(k_\ell)}^\dagger\,,
\end{equation}
 because in both cases the dissipative generator $\Lambda$ commutes with the precession.

As initial state we use an SU(2) coherent state, which can be seen as the most
classical state of a spin 
\cite{Giraud08,Giraud10}, and is usually easy to prepare (for instance by optically polarizing the atomic spins in a SERF magnetometer). Also, it is equivalent to 
a symmetric state of $2j$ spin-$\frac{1}{2}$ pointing all in the same direction. With respect to the  $\ket{j,m}$ basis it reads 
\begin{align}
\ket{j,\theta,\phi} =\sum_{m=-j}^j \sqrt{\binom{2j}{j-m}} \sin \left(\frac{\theta}{2}\right)^{j-m} \cos \left(\frac{\theta}{2}\right)^{j+m}e^{i(j-m)\phi} \ket{j,m}\,.
\label{spin coherent states}
\end{align}
We choose $\theta=\frac{\pi}{2}$, $\phi=\frac{\pi}{2}$.\\

\lu{\section{Optimizing the kicked top}}

\subsection{The kicked top as a control problem}\label{kt_as_control_problem}

We consider the kicked top as a control problem and discretize the kicking strengths $k_\ell$ and times $t_\ell$. The precise parameters of the discretized control problem vary between the following examples and are summarized in Appendix \ref{app:control_problem_params}.  In the following, $t_\text{step}$ denotes a discrete time step (measured in units of $\tau =1$), $k_\text{step}$ is a discrete step of kicking strength, the RL agent optimizes the QFI at time $T_\text{opt}$, and we bound the total accumulated kicking strength $\sum_{\ell}k_\ell<15000$ which is never reached in optimized policies though. The frequency $\omega$, that we want to estimate, is set to induce a rotation of the state by $t \pi/2$ ($t$ is measured in units of $\tau=1$).

Possible control policies are simply given by a vector of kicking strengths $\boldsymbol{k}=(k_1,\dotsc,k_N)\in\mathbb{R}^N$ with $k_\ell\in \{qk_\text{step}: q=0,1,2,\dotsc\}$. To each policy corresponds a QFI value, calculated from the resulting state $\rho (T_\text{opt})$, which quantifies how well the policy performs. To tackle this type of problem, various numerical algorithms are available, each with its own advantages and drawbacks \cite{dunjko2018machine, mehta2019high, palittapongarnpim2017learning}. We pursue the relatively unexplored (in the context of physics) route of cross-entropy RL.

\subsection{ Reinforcement learning}
\begin{figure}[h!]
	\centering
	\includegraphics[width=0.7\linewidth]{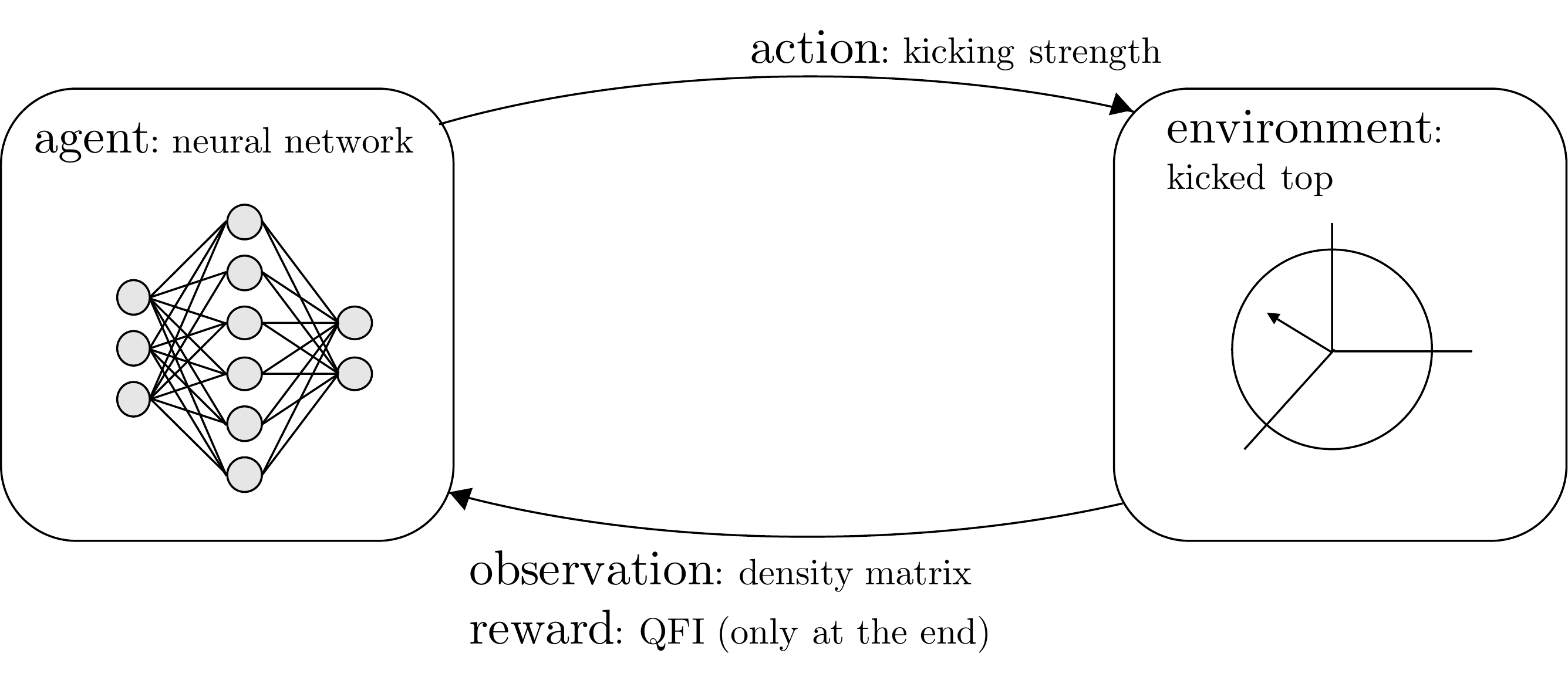}
	\caption{Typical setup of reinforcement learning: the RL agent acts upon the environment
		which in return gives the RL agent an observation and a reward. In our case the RL agent is a neural network and the environment is the generalized kicked top.}
	\label{fig:RL}
\end{figure}

\lu{Fig.~\ref{fig:RL} shows the typical way we imagine RL. There is an \textit{agent} that interacts with an \textit{environment} by choosing \textit{actions} and receiving an \textit{observation }and a \textit{reward} from the environment. One cycle of action and observation/reward is called a \textit{step}.}

\lu{In general, the idea of reinforcement learning is to reinforce behaviour that leads to high rewards. The precise mechanism depends on the used RL algorithm. }

\subsection{The kicked top as a reinforcement learning problem}

\lu{The system, the generalized kicked top as introduced in Section \ref{the_system}, represents the RL environment.
The agent can choose between only two actions: (i) increase the kicking strength (by $k_\text{step}$) or (ii) go on from the current position in time $\ell t_\text{step}$ to $(\ell+1)t_\text{step}$. In this way, the vector $\boldsymbol{k}$ is built up step by step. After each action, the agent obtains an observation given by the full density matrix of the current state of the environment. Since we simulate the evolution of the environment, the density matrix is readily available.

Only after the total time $T_\text{opt}$, a reward [the QFI of $\rho(T_\text{opt})$] is given to the agent. This concludes one \textit{episode}, and the resulting vector $\boldsymbol{k}$ represents a policy.} Then, the environment is reset [i.e., the spin is reinitialized with the coherent state at $\theta=\frac{\pi}{2}$, $\phi=\frac{\pi}{2}$, see Eq.~(\ref{spin coherent states})], and the next episode starts.

A neural network represents the RL agent: 
The observation is given to the neural network's input neurons while each output neuron represents one possible action, i.e., we have two output neurons for ``kick'' and ``go on''.  The activation of these output neurons determines the probability of executing that action. The policy, however, is not given by the neural network directly. Since the environment is deterministic (i.e., the state evolves deterministically for a given policy $\boldsymbol{k}$ of kicking strengths) there is no point in choosing a stochastic policy such as a neural network. Instead, a single choice of kicking strengths $\boldsymbol{k}$ represents the policy. We obtain this by first training the neural networks using the cross-entropy method, then generating a few episodes with the trained neural network, and then picking the episode with the largest QFI. The kicking strengths applied in that episode represent the policy\footnote{In comparison, Sanders et al.~\cite{hentschel2010machine, hentschel2011efficient, lovett2013differential} restricted their policy search for adaptive single-photon interferometry in such a way that their search space corresponds to points in $\mathbb{R}^N$, making it similar to our problem. However, in their case the observations from the environment are probabilistic measurement outcomes while in our case the observation is the deterministic state $\rho$.}.

\subsection{Cross-entropy method}

\lu{The RL cross-entropy method \cite{de2005tutorial} we use works as follows: We first produce a set of episodes (i.e., we obtain several vectors $\boldsymbol{k}$) with a neural network that is initialized randomly. Then, we rank those episodes according to their reward\footnote{\lu{We do not give an immediate reward at every step but only at the very end of an episode, and the reward is not reduced with the number of steps (i.e., the discount factor is 1).} }. We select the best $10\% $ of episodes (with highest reward) for further computations. Every episode can be split into several pairs of action and observation and we use those pairs to train the neural network with the stochastic gradient descent method called \textit{Adam} \cite{kingma2014adam}. 
As a result of this training, the weights of the neural network are adjusted, i.e., the agent learns from its experience. Future actions taken by the agent are influenced not only by randomness but also by this experience. One run of producing episodes, ranking them, and using the best $10\% $ to train the neural network is called an \textit{iteration}. Training a neural network consists of several iterations. See Appendix \ref{app:pseudocode} for pseudocode of this algorithm.}
For the parameters of the training process see Appendix \ref{app:RL_params}. In Appendix \ref{app:learning_curve_and_stability} we study the learning success for different numbers of episodes and iterations.

\section{Results}
\begin{figure}[h!]
	\centering
	\includegraphics[width=0.8\linewidth]{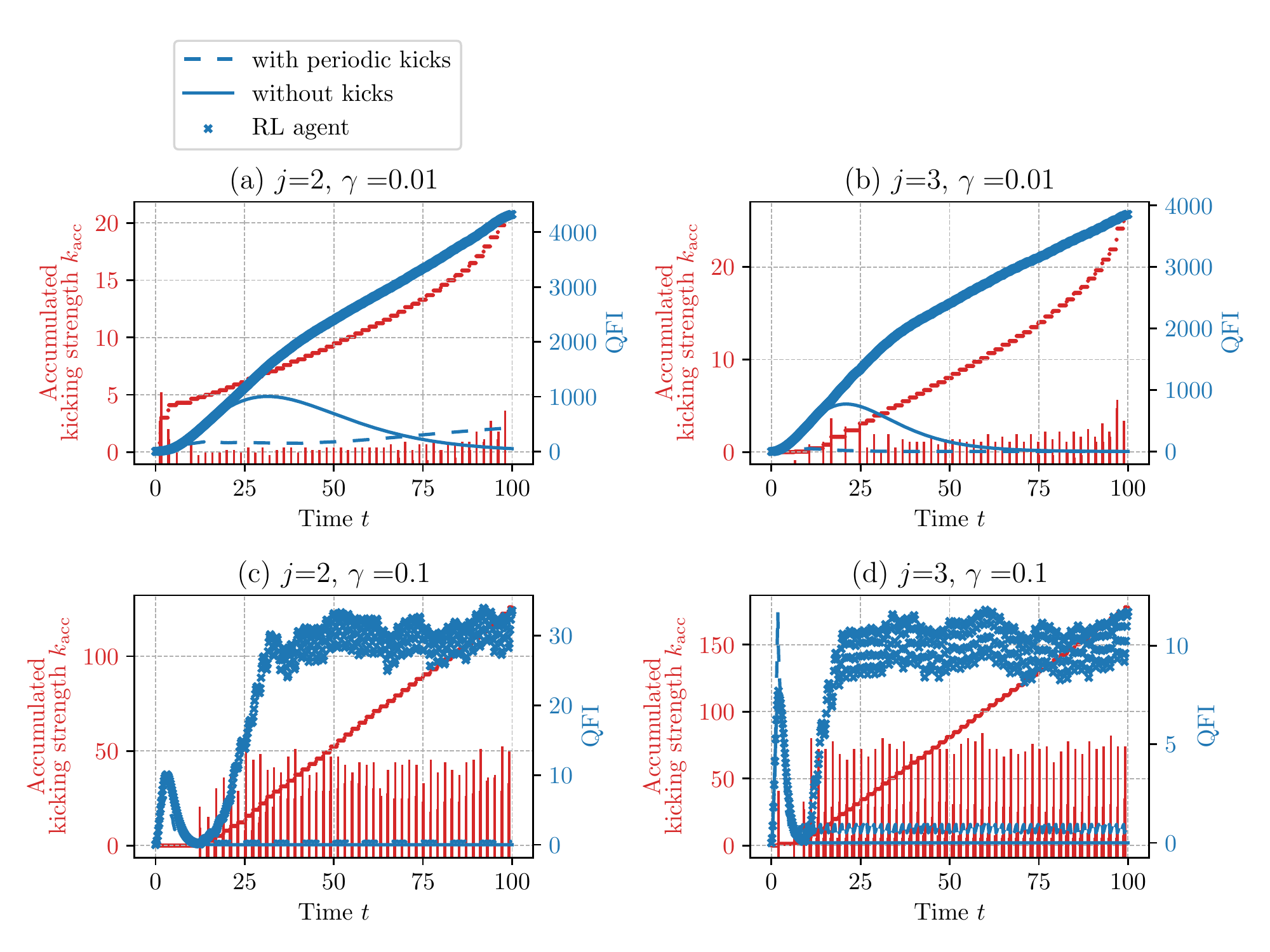}
	\caption{Examples for the policy adopted by the RL agent
		for superradiant damping.
		We plot the accumulated kicking strength \lu{$k_\text{acc}(t)$ (integrating over the kicks from time $0$ to $t$)} on the left axes as red dots and on the right axes in blue the quantum Fisher information for the top (solid line), the periodically kicked top with $k=30$ chosen as in Ref.~\cite{fiderer2018quantum} (dashed line) and the QFI that corresponds to the policy of the RL agent (crosses). We additionally plot red vertical lines in the places, where the RL agent decides to set a kick. The height of the lines correspond to the kicking strength in arbitrary units and are not on the scale of the left axis.
		There is a regime where the RL agent manages to increase the QFI with each time step [panel (a) and (b)], and a regime where the RL agent 
		makes the QFI oscillate [panel (c) and (d)].
	}\label{fig:superradiant_strategies}
\end{figure}

We compare the QFI for different models: (i) the top (simple precession without kicks), (ii) the standard kicked top, as studied in Ref.~\cite{fiderer2018quantum}, with periodic kicks (period $\tau=1$, i.e., a precession angle of $\pi/2$ for one period, and kicking strength $k=30$), and (iii) the generalized kicked top optimized with RL. 
In case of superradiance damping (phase damping) we denote the top by SR-T (PD-T), the standard kicked top by SR-KT (PD-KT) and the RL-optimized generalized kicked top by SR-GKT (PD-GKT). Details on the training and the optimization of the RL results are provided in Appendix \ref{app:RL_params}.

 Let us first consider superradiant damping with results presented in Fig.~\ref{fig:superradiant_strategies}. The QFI for the SR-T exhibits a characteristic growth quadratic in time. However, due to decoherence, the QFI does not maintain this growth but starts to decay rapidly towards zero. The time when the QFI reaches its maximum was found to decay roughly as $1/(\gamma_\text{sr} j)$ with spin size $j$ and damping rate $\gamma_\text{sr}$ \cite{fiderer2018quantum}.
 
The situation changes with the introduction of nonlinear kicks.
There, the QFI for the SR-KT shows the interesting behavior of not decaying to zero for large times. Instead it reaches a plateau value which was found to take surprisingly high values for specific choices of $j$ and dissipation rates \cite{fiderer2018quantum}, in particular, for $j=2$. The system looses energy through superradiant damping while the nonlinear kicks add energy. This prevents the state from decaying to the ground state, which is an eigenstate of the precession and would lead to a vanishing QFI. From this perspective, the plateau results from a dynamical equilibrium established by the interplay of superradiant damping and kicks.

However, the full potential of exploiting such effects and increasing the QFI with the help of nonlinear kicks is not achieved with constant periodic kicks. Instead, the RL agent\footnote{The training of one RL agent takes about eight hours on a desktop computer.} finds policies to make the QFI of the SR-GKT increase further even when the QFI of the SR-T decayed already to zero and the QFI of the SR-KT reached its plateau value.

Examples for $j=2$ and $j=3$ are presented in Fig.~\ref{fig:superradiant_strategies}. The QFI of the SR-GKT is optimized for a total time $T_\text{opt}$ which is the largest time plotted in each example. At $T_\text{opt}$, the plateau value of the SR-KT for $j=3$ is relatively low and the RL-optimized policy achieves an improvement in sensitivity (associated with $1/\sqrt{I_\omega}$) of more than an order of magnitude.
Panels (a) and (b) show continuous growth of the QFI through an optimized kicking policy. 
Only if the time $T_\text{opt}$ (the QFI is optimized to be maximal at $T_\text{opt}$) is increased further, the impressive growth of the QFI finally breaks down. Instead of increasing $T_\text{opt}$, we choose to increase superradiant damping while keeping $T_\text{opt}$ constant, which has a similar effect. In that case, see panels (c) and (d), the RL agent chooses a policy which makes the QFI oscillate at a relatively high level before the time $T_\text{opt}$ is reached.

\begin{figure}[htbp!]
	\centering
	\includegraphics[width=0.8\linewidth]{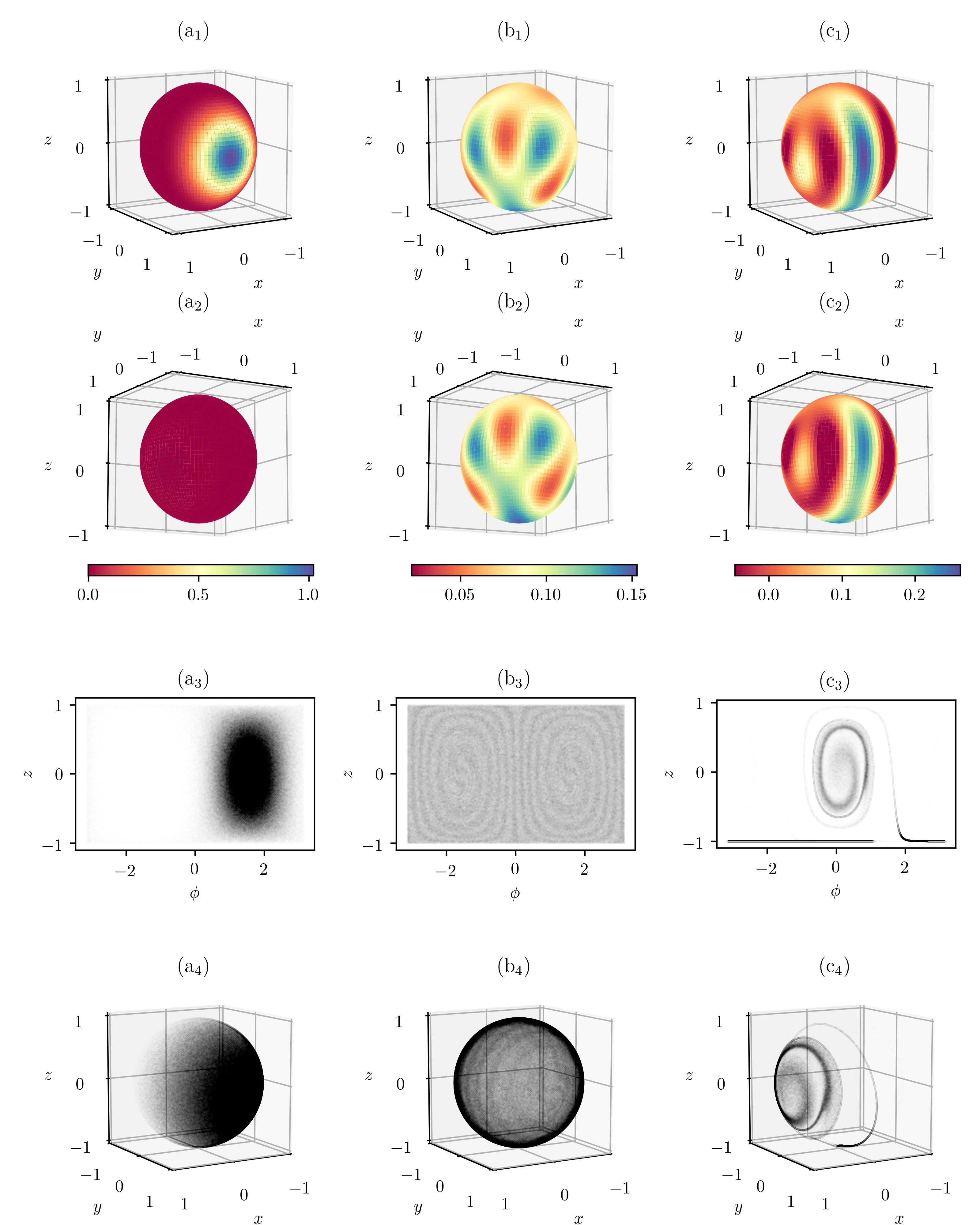}
	\caption{Illustration of kicked superradiant dynamics with Wigner functions and its classical limit. The spin size is $j=3$ and the dissipation rate is $\gamma_\text{sr}=0.01$. Panels in the left column (a) correspond to the initial spin coherent state at $\theta=\phi=\pi/2$. The middle and right columns correspond to the state at time $T_\text{opt}$ generated with periodic kicks (middle column (b), $k=30$) and with kicks optimized with reinforcement learning [right column (c), the corresponding QFI is shown in panel (b) of Fig.~\ref{fig:superradiant_strategies}].  The top two rows show the Wigner functions of the density matrix, the bottom two rows show the classical phase space, populated by $10^6$ points initially distributed according to the Husimi distribution of the initial spin coherent state and then propagated according to the classical equations of motion.}
	\label{fig:classical_and_wigner}
\end{figure}

The superiority of the policies found by the RL agent can be understood by taking a look at the evolution of the quantum state, see Fig.~\ref{fig:classical_and_wigner}: We represent the quantum state in the space of $\boldsymbol{r}=(x,y,z)=(\braket{J_x},\braket{J_y},\braket{J_z})$ where $\braket{J_\ell}\coloneqq \text{tr}(\rho J_\ell)$ and, due to the conservation of angular momentum, $|\boldsymbol{r}|=1$ which restricts the space to a sphere. This is represented in Fig.~\ref{fig:classical_and_wigner} with either a sphere parametrized with $x$, $y$, and $z$, or in a plane (the phase space) spanned by the $z$-coordinate and the azimuthal angle $\phi\in(-\pi,\pi]$ such that $\phi=z=0$ corresponds to the positive $x$-axis, $\phi=\pi/2$, $z=0$ to the positive $y$-axis, and $z=\pm 1$ with arbitrary $\phi$ to the positive (negative) $z$-axis.

The quantum state can be represented in the phase space with the help of the Husimi or
the Wigner distributions which are quasi probability distributions of the quantum state. 
The first two rows of panels in Fig.~\ref{fig:classical_and_wigner} depict the Wigner distribution of the initial quantum state (left column) and the quantum states of the SR-KT (middle column, with kicking strength $k=30$) and SR-GKT (right column) evolved for a time $T_\text{opt}$ with damping rate $\gamma_\text{sr}=0.01$. The plotted cases for  the SR-KT and  SR-GKT correspond to the QFI given in panel (b) of Fig.~\ref{fig:superradiant_strategies}, where one can also see the corresponding RL-optimized distribution of kicks.
 
Due to the small spin size of $j=3$, we are deep in the quantum mechanical regime which manifests itself in an uncertainty of the initial spin coherent state that is relatively large compared to total size of the phase space. The distribution of the states evolved under dissipative dynamics exhibit remarkable differences for periodic and RL-optimized kicks: 

In case of periodic kicks, we find that the initially localized distribution gets distributed over the phase space. It exhibits a maximum on the negative $z$-axis, see panels (b$_1$) and (b$_2$)in Fig.~\ref{fig:classical_and_wigner}. This is reminiscent of the dissipative evolution in the absence of kicks, where the state is driven towards the ground state $\ket{j,-j}$ which is centered around $z=-1$. The ground state $\ket{j,-j}$ is an eigenstate of the precession and, thus, insensitive to changes in the frequency $\omega$ we want to estimate. Similarly, we interpret the part of the state distribution of the SR-KT that is centered around negative $z$-axis as insensitive.  However, the distribution also exhibits non-vanishing parts distributed over the remainder of the phase space that can be understood as being sensitive to changes of $\omega$ and therefore explain the non-zero QFI of the SR-KT.

The state corresponding to RL-optimized kicks looks like a strongly squeezed state that almost encircles the whole sphere. Similar to spin squeezing, which is typically applied to the initial state as a part of the state preparation, we interpret the squeezed distribution as particularly sensitive with respect to the precession dynamics. This is due to the reduced uncertainty along the precession trajectories, i.e., with respect to the $\phi$ coordinate. We provide clips of the evolution over time of the state distributions that illustrate how the RL agent generates the squeezed state\footnote{The clips are available at \href{https://doi.org/10.6084/m9.figshare.c.4640051.v3}{https://doi.org/10.6084/m9.figshare.c.4640051.v3}.}. In particular, the squeezed state distribution can be seen as a feature the RL agent is aiming for with its policy. The distribution of RL-optimized kicks is shown in Fig.~\ref{fig:superradiant_strategies} (in Appendix \ref{app:distribution_of_kicks}, we provide a finer resolution of the distribution of kicks): It is roughly periodic with period corresponding to a precession angle of $\pi$. Also note that for the SR-GKT the Wigner distribution has negative contributions which is associated with non-classicality of the quantum state \cite{agarwal2012quantum}.

An advantage of the superradiant dynamics lies in its well-defined simple classical limit \cite{Braun01B}, see also Appendix \ref{app:classical_equations}. The lower two rows of panels in Fig.~\ref{fig:classical_and_wigner} depict the corresponding classical limit where the quantum state is represented by a cloud of phase space points (distributed according to the Husimi distribution of the initial spin coherent state) that are propagated according to the classical equations of motion. One of the reasons why the evolved classical distributions differ from the Wigner distributions is the absence of quantum uncertainty in the classical dynamics; in principle, over the course of the dynamics all classical phase space points can be concentrated to an arbitrarily small region of the phase space. In case of the SR-KT, the phase space points are distributed over the whole phase space, reminiscent of classical chaos. However, the distribution is not completely uniform but it exhibits a spiral density inhomogeneity. 
The plots as in Fig.~\ref{fig:classical_and_wigner} but for $j=2$ are shown in the Appendix \ref{app:distribution_of_kicks}.

\begin{figure}[h!]
	\centering
	\includegraphics[width=0.7\linewidth]{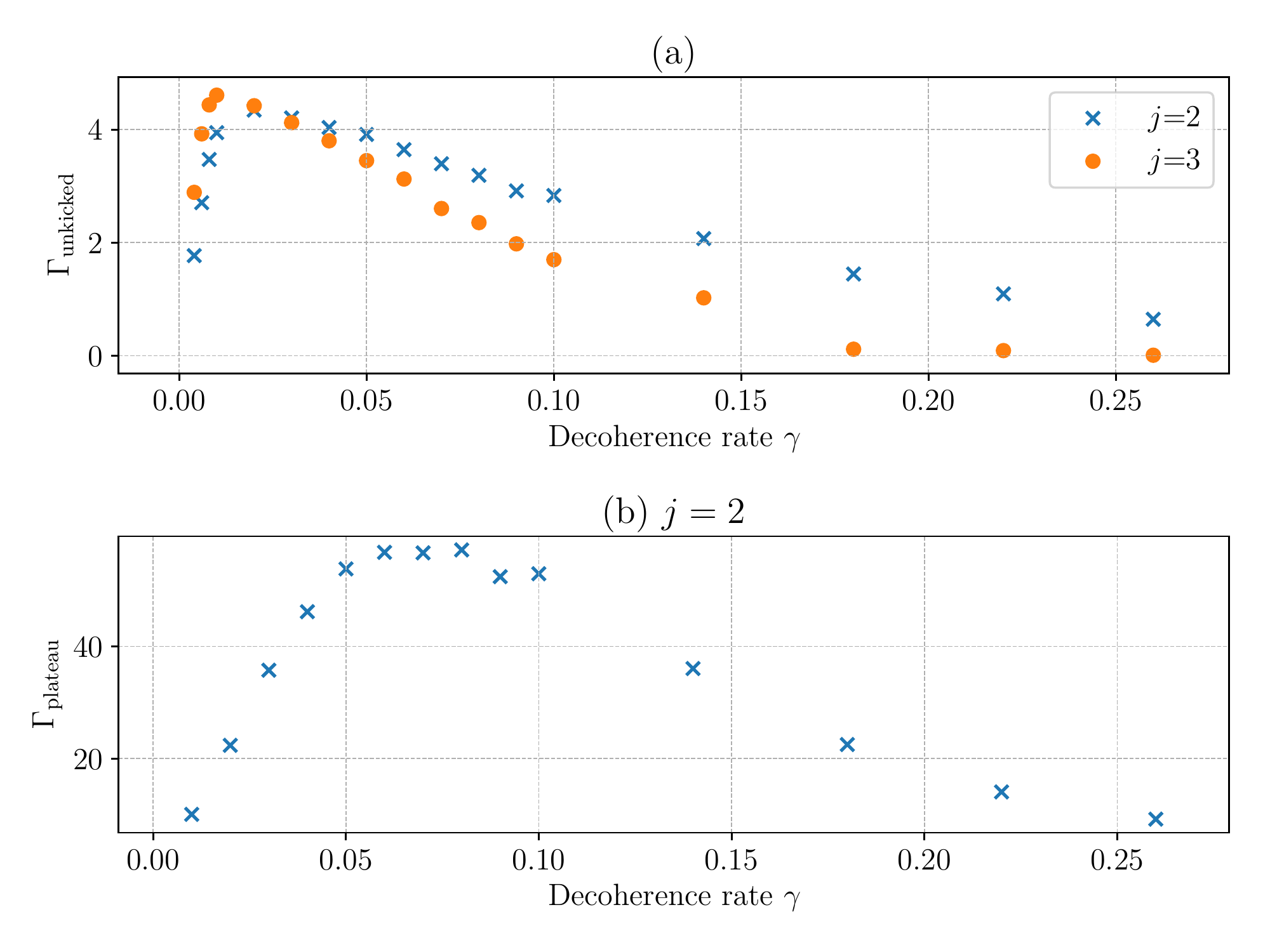}
	\caption{Improvement in the quantum Fisher information due to reinforcement learning for superradiant damping.
		The improvement in panel (a) is the ratio \lu{$\Gamma_\text{unkicked}$} of quantum Fisher information at time $T_\text{opt}$ (100 discretized time steps) optimized with reinforcement learning and the maximum QFI of the top (no kicks). In panel (b) we plot the ratio \lu{$\Gamma_\text{plateau}$} of the QFI optimized with reinforcement learning and the plateau values achieved by periodic kicking for spin size $j=2$ and kicking strength $k=30$. In panel (b), the case of $j=3$ is omitted due to the very small plateau values in that case. The discretization is coarser than in previous examples: $t_\text{step}=1$ (i.e., a precession angle of $\pi/2$ per time step) and $k_\text{step}=0.1$.}
	\label{fig:periodic_kicking_gain}
\end{figure}
Fig.~\ref{fig:periodic_kicking_gain} shows the gains of the RL-optimized SR-GKT over the SR-T.
The gain is defined as the ratio of the RL-optimized QFI at time $T_\text{opt}$ and the maximum QFI for the  SR-T. A broad damping regime is found where gains can be achieved:
In the regime of small decoherence rates $\gamma_\text{sr}$, the RL agent can fight decoherence in such a way that the QFI exhibits a continuous growth over the total time $T_\text{opt}$ [see panels (a) and (b) in Fig.~\ref{fig:superradiant_strategies}]. In comparison with the SR-T, the RL agent benefits of stronger damping in this regime and, therefore, the gain increases with the dissipation rate $\gamma_\text{sr}$.  For larger decoherence rates, the RL agent can no longer fight decoherence in the same manner [see panels (c) and (d) in Fig.~\ref{fig:superradiant_strategies}], which manifests itself in a reduction of gains for large decoherence rates.
In panel (b) of Fig.~\ref{fig:periodic_kicking_gain}, we can see the (even larger) gain in QFI compared to the plateau value reached by the SR-KT.

\begin{figure}[htbp!]
	\centering
	\includegraphics[width=0.8\linewidth]{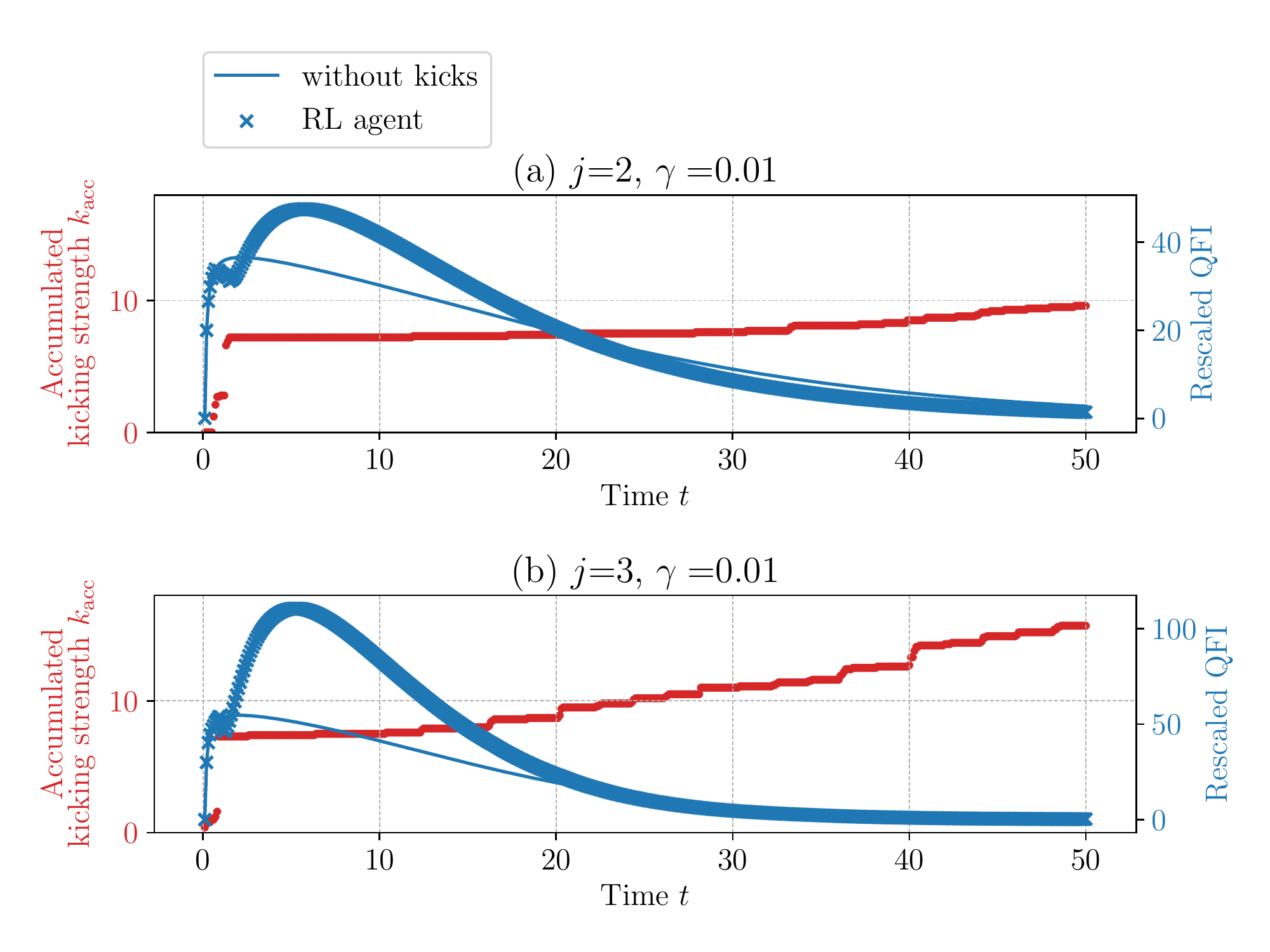}
	\caption{Examples for the policy adopted by the RL agent for maximizing the rescaled quantum Fisher information with superradiant damping. We plot the accumulated kicking strength   \lu{$k_\text{acc}(t)$ (integrating over the kicks from time $0$ to $t$)} on the left axis as red dots and on the right axis the rescaled quantum Fisher information for the top (blue solid line) and for the generalized kicked top optimized with reinforcement learning (blue crosses). In case of $j=2$ ($j=3$) the strongest kick is applied after an initial rotation angle of $13\pi/20$ ($9\pi/20$). }
	\label{fig:rescaled_qfi}
\end{figure}
\begin{figure}[htbp!]
	\centering
	\includegraphics[width=0.8\linewidth]{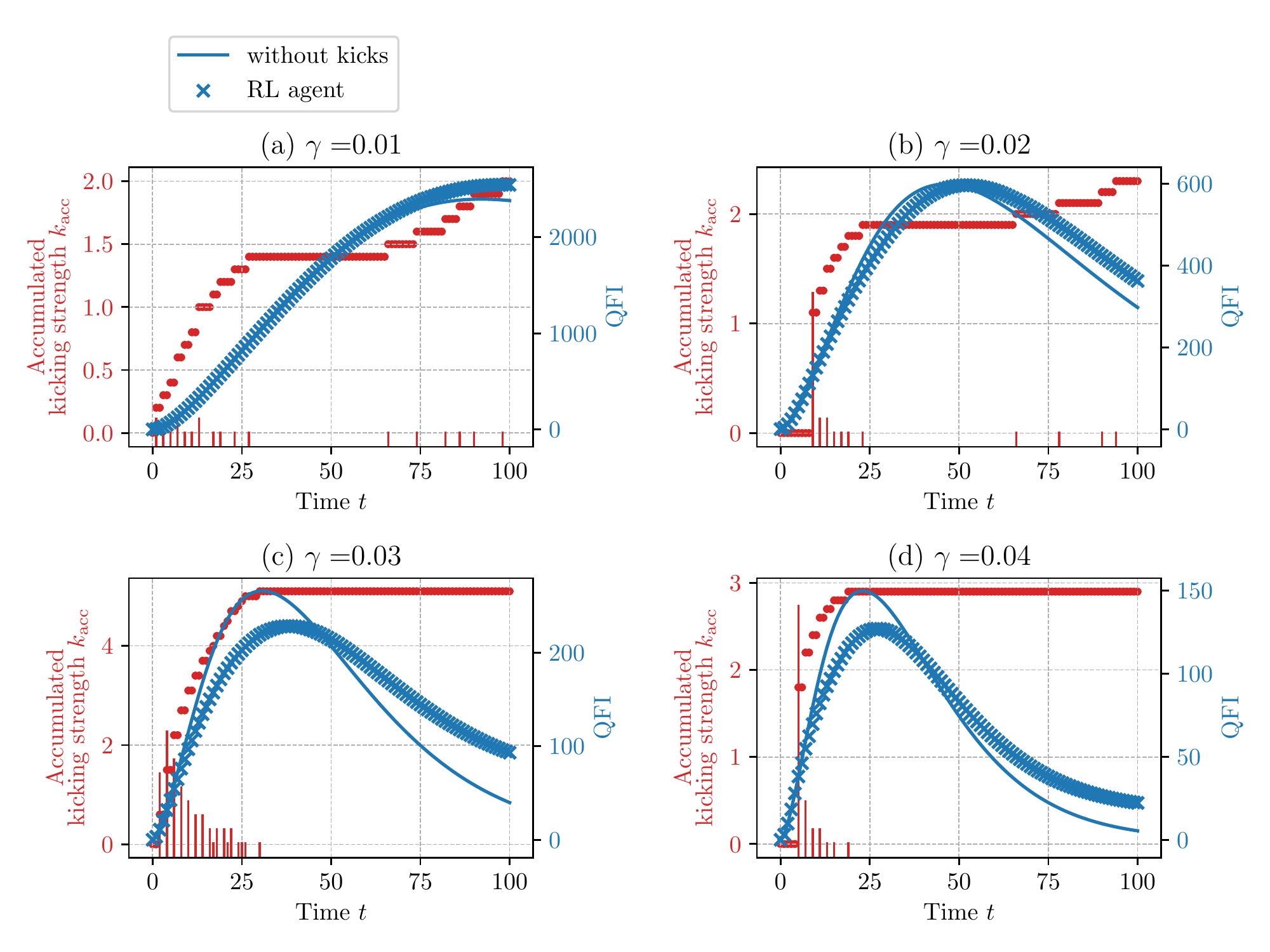}
	\caption{Examples for the strategy adopted by the RL agent for phase damping. All data is for spin $j=2$ with increasing damping rates $\gamma_\text{pd}$ from panel (a) to (d). We plot the accumulated kicking strength \lu{$k_\text{acc}(t)$ (integrating over the kicks from time $0$ to $t$)} on the left axis as red dots and on the right axis the quantum Fisher information for the top (blue solid line) and for the generalized kicked top optimized with reinforcement learning (blue crosses). We additionally plot red vertical lines at times when the RL agent sets a kick. The length of the lines corresponds to the kicking strength in arbitrary units (independent of the scale of the left axis). Note that the RL agent aims to maximize the QFI for $T_\text{opt}=100$ and outperforms the top in all examples.}
	\label{fig:phasedamping_strategies}
\end{figure}

The RL-optimized QFI is associated with a lower bound on the sensitivity (see Eq.~\ref{eq:cramer_rao}) for a given measurement time $T_\text{opt}$. If measurement time can be chosen arbitrarily, sensitivity is associated with $\max_t I_\omega(t)/t$ \cite{fiderer2018quantum}. This sensitivity represents the standard quantity reported for experimental parameter estimation because it takes time into account as a valuable resource; sensitivity is given in units of the parameter to be estimated per square root of Hertz. With RL we try to maximize $\max_t I_\omega(t)/t$ with respect to policies.

Fig.~\ref{fig:rescaled_qfi} compares the SR-T with the SR-GKT where the latter was optimized with RL in order to maximize the rescaled QFI. Note, that the initial spin coherent state is centered around the positive $y$-axis, which means it is an eigenstate of the nonlinear kicks; kicks cannot induce spin squeezing at the very beginning of the dynamics. This changes when the spin precesses away from the $y$ axis. Therefore, it makes sense that the RL agent applies the strongest kick  only after a precession by about $\pi/2$. 
The actions that the RL agent takes after the rescaled QFI reached its maximum are irrelevant and can be attributed to random noise generated by the RL algorithm.

As we have seen, the interplay of nonlinear kicks and superradiant damping is very special. However, also for other decoherence models the QFI can be increased significantly, for instance in case of a alkali-vapor magnetometer \cite{fiderer2018quantum}. To demonstrate the performance of the RL agent in connection with another decoherence model, we take a look at phase damping, see
Fig.~\ref{fig:phasedamping_strategies}. The behavior of the QFI of the PD-T is qualitatively similar to superradiant damping.
The introduction of kicks, however, has a qualitatively different effect on the QFI. The RL agent can achieve improvements of the QFI for the PD-GKT at time $T_\text{opt}$ (the highest time plotted in each panel of Fig.~\ref{fig:phasedamping_strategies}) compared with the QFI of the PD-T at the same time. Compared to the superradiant case, improvements are rather small. Notably, the policies applied by the RL agent are also different from superradiant damping; for instance, the RL agent avoids kicks for large parts of the dynamics.

\section{Discussion}

This work builds on recent results on quantum-chaotic sensors \cite{fiderer2018quantum}. \lu{Our aim is to optimize the dynamical control that was used in Ref.~\cite{fiderer2018quantum} to render the sensor dynamics chaotic. Due to the high dimensionality of the problem we use techniques from reinforcement learning (RL).} The control policies found with RL are tailored to boundary conditions such as the initial state, the targeted measurement time, and the decoherence  model under consideration. At the example of superradiant damping we demonstrate improvements in measurement precision and an improved robustness with respect to decoherence. A drawback of RL often lies in the expensive hyperparameter tuning of the algorithm. However, here we show that a basic RL algorithm (the cross-entropy method) can be used for several choices of boundary conditions with practically no hyperparameter tuning (there was no hyperparameter search necessary, solely parameters that directly influence the computation time were chosen conveniently).
 
In the example of superradiant damping, we unveil the approach taken by RL by visualizing the quantum dynamics with the help of the Wigner distribution of the quantum state. This reveals that RL favors a policy  that is reminiscent of spin squeezing. However, instead of squeezing the state only at the beginning of the dynamics, the squeezing is refreshed and enhanced in roughly periodic cycles in order to fight against the superradiant damping.

In the spirit of Ref.~\cite{fiderer2018quantum}, these findings emphasize the potential that lies in the optimization of the measurement dynamics. We are optimistic that reinforcement learning \lu{can be used to tackle other problems} in quantum metrological settings in order to achieve maximum measurement precision with limited quantum resources.

\section*{Acknowledgments}
 L. J. F. and D. B. acknowledge support from the Deutsche Forschungsgemeinschaft (DFG), Grant No. BR 5221/1-1.
 
\clearpage
\appendix 
\section{Control problem and optimisation parameters of the examples}\label{app:control_problem_params}
 Table \ref{tab:parameters} shows the parameters of the control problem and for the optimization used in each example. We train $n_\text{agents}$ RL agents for $n_\text{iterations}$ iterations with  $n_\text{episodes}$ episodes in each iteration. Each episode is simulated until a total time $T_\text{opt}$ is reached. Then we produce $n_\text{samples}$ sample episodes of each trained RL agent and choose the best episode to plot the sample policies and gains.
\begin{table}[htbp!]
\caption{Hyperparameters used for the examples in the main text.}
\begin{tabular}{l ||c|c|c|c|c|c|c}
	 Figure & $n_\text{agents}$ & $n_\text{iterations}$ & $n_\text{episodes}$ & $n_\text{samples}$ & $t_\text{step}$ & $k_\text{step}$ & $T_\text{opt}$ \\ 
	\hline\hline Samples with superradiant damping (Fig.~\ref{fig:superradiant_strategies}) & 5 & 500 & 50 & 20 & 0.2 & 0.05 &  100\\ 
	\hline  Gains of superradiant damping (Fig.~\ref{fig:periodic_kicking_gain})  & 20 & 300 & 40 & 20 & 1.0 & 0.10 & 100 \\ 
	\hline Samples of rescaled QFI (Fig.~\ref{fig:rescaled_qfi})   & 2 & 500 & 50 & 20 & 0.1 & 0.10 &  50\\ 
	\hline Samples with phase damping (Fig.~\ref{fig:phasedamping_strategies}) & 1 & 1,000 & 100 & 1 & 1.0 & 0.10 & 100 \\ 
\end{tabular} 
\label{tab:parameters}
\end{table}

\lu{\section{Hyperparameters of reinforcement learning}} \label{app:RL_params}

Here we give further information on the neural network \lu{and the hyperparameters of the algorithm.}

The input layer of the neural network is defined by the observation. The output layer is determined by the number of actions (two) and we choose 300 neurons in the hidden layer. The layers are fully connected. The hidden layer has the rectified linear unit (ReLU) as its activation function and the output layer has the softmax function as its activation function \cite{nielsen2015neural}. As a cost function we choose the categorical cross entropy \cite{nielsen2015neural}. The share of best episodes $\sigma_\text{share}$ is always $10\%$. The number of iterations and number of episodes vary for different settings, see Table \ref{tab:parameters} for detailed information. For training we use the Adam optimizer \cite{kingma2014adam} with learning rate $0.001$.

\section{Pseudocode for cross-entropy reinforcement learning} \label{app:pseudocode}

\lu{This is the pseudocode for the cross-entropy method with discrete actions.}

\begin{figure}[h!]
	\centering
	\includegraphics[width=1.0\linewidth]{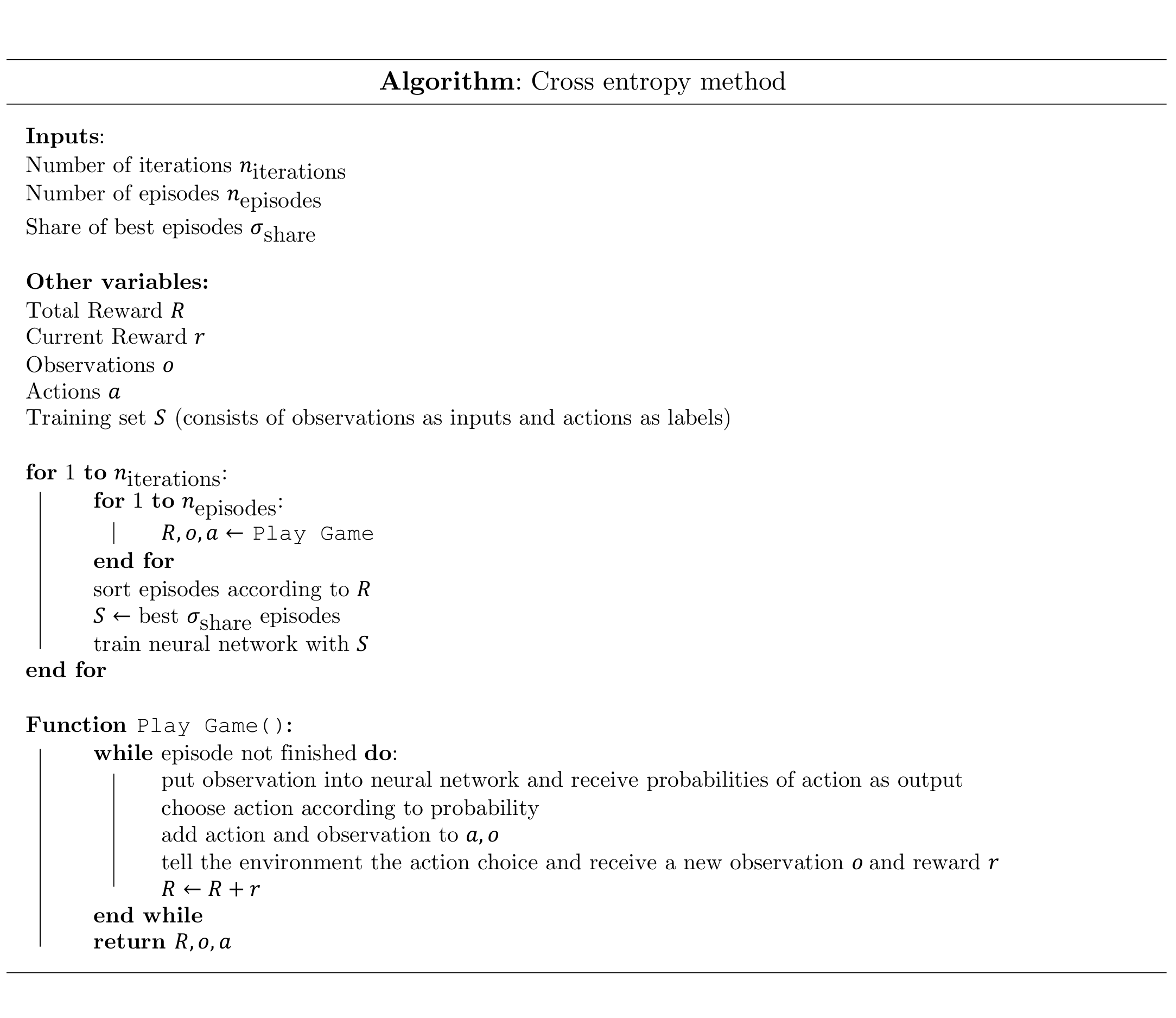}
\end{figure}

The code implementation is based on an example by Jan Schaffranek \footnote{https://www.udemy.com/artificial-intelligence-und-reinforcement-learning-in-python}. 
 
\section{Learning curve and stability of the algorithm} \label{app:learning_curve_and_stability}
\begin{figure}[h!]
	\centering
	\includegraphics[width=0.7\linewidth]{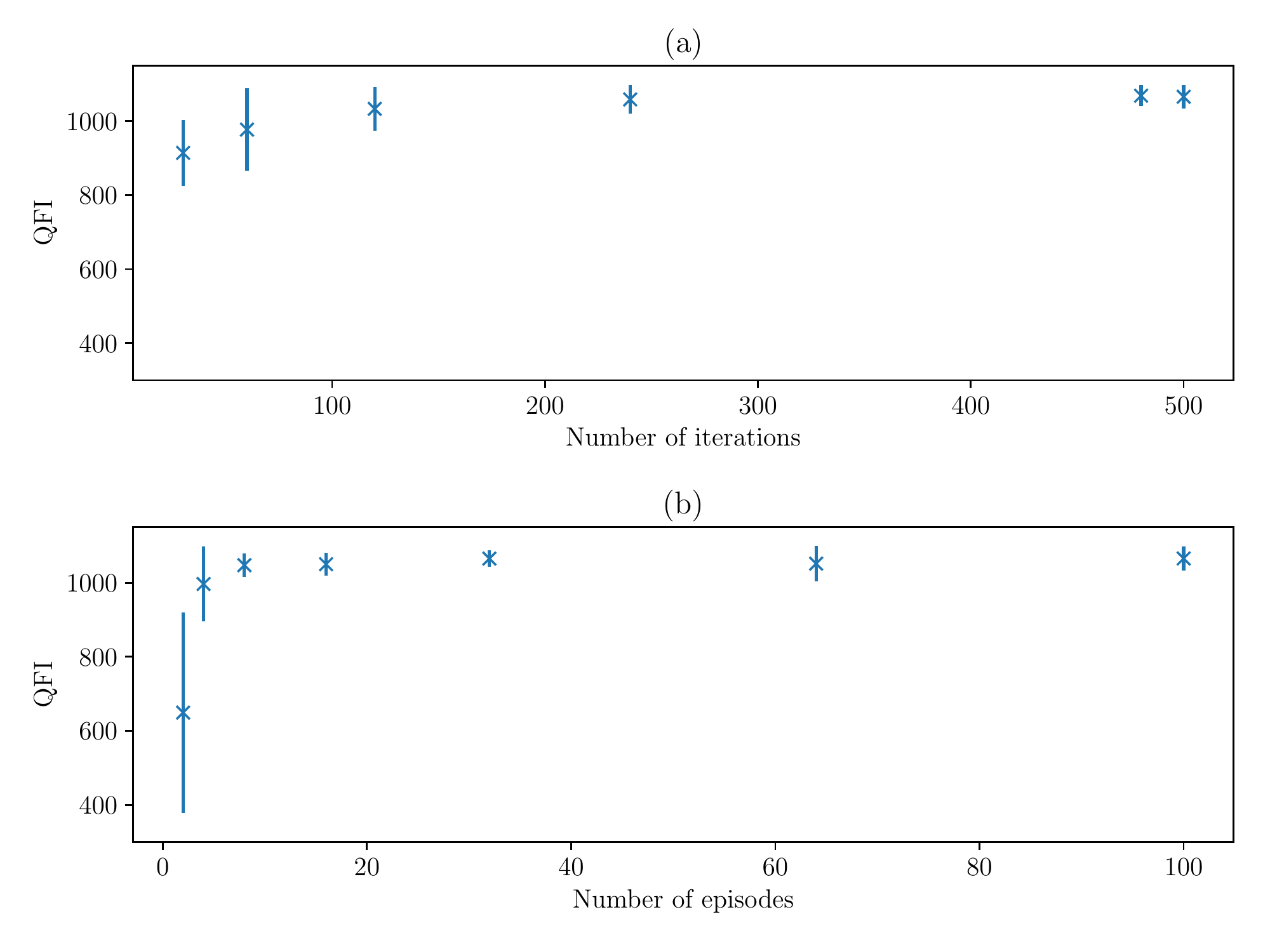}
	\caption{Learning behavior of the algorithm at the example of superradiant damping with $j=2$, $\gamma_\text{sr}=0.02$, $k_\text{step}=0.1$, $t_\text{step}=1$, and $T_\text{opt}=100$. Panel (a) shows how the mean QFI and its standard deviation with respect to different runs of the algorithm behaves for various numbers of iterations and fixed number of episodes fixed to 100. In Panel (b)  the number of episodes is varied and number of iterations fixed to 500.}
	\label{fig:learning_rates}
\end{figure}
At the example of the superradiance decoherence model, we study the learning behavior of the cross-entropy reinforcement learning algorithm for different training lengths (i.e.~number of iterations) and different numbers of episodes per iteration. 
The results are summarized in Fig.~\ref{fig:learning_rates}. Spin size is  $j=2$ and dissipation rate is $\gamma_\text{sr}=0.02$. 

In order to see the influence of the number of iterations, we set the number of episodes to 100 and let 20 different RL agents (with different random seeds) train for various numbers of iterations. The training of a single RL agent takes about one hour at most (for the higher number of iterations) on a desktop computer.
We then use each RL agent to produce 20 episodes, giving us 400 episodes for each data point in Fig.~\ref{fig:learning_rates}. We used those episodes to calculate mean and standard deviation of the reward.
 The results are shown in the panel (a) of Fig.~\ref{fig:learning_rates}. In order to see the influence of the number of episode in each iteration, we fix the number of iterations to 500 and do the same procedure as before. 
 The results are shown in panel (b) of Fig.~\ref{fig:learning_rates}.

We can see that the standard deviation over policies decreases with the number of iterations while the mean QFI increases. The same is true for the number of episodes [panel (b)], where for 32 episodes a stable plateau of the QFI is reached such that increasing the number of episodes does not achieve any further improvements. Overall, these results demonstrate the stability of the algorithm if the number of episodes and iterations is chosen sufficiently large.
 \section{Classical equations of motion}\label{app:classical_equations}
 The kicked top with superradiant damping has a well defined classical limit. It is obtained from the quantum equations of motion by taking the limit $j\rightarrow \infty$ where $\hbar=1$ and $\tau=1$. The rescaled angular momentum operator $2\boldsymbol{J}/(2j+1)=2(J_x,J_y,J_z)/(2j+1)$ then becomes the classical coordinate vector $\boldsymbol{r}=(x,y,z)$ and with $\lim_{j\rightarrow\infty}\left(\frac{2\boldsymbol{J}}{2j+1}\right)^2=1$ the unit sphere becomes the classical phase space with azimuthal angle $\phi$ and $z$-coordinate as canonical variables. The equations of motions $\boldsymbol{r}\rightarrow \tilde{\boldsymbol{r}} $ are found to be \cite{Braun01B}
 \begin{align}
 \tilde{x}&=x \cos(\alpha)-y \sin(\alpha),\\
 \tilde{y}&= x \sin(\alpha)+y \cos(\alpha),\\
 \tilde{z}&=z,
 \end{align}
 for the precession about the $z$-axis by an angle $\alpha$,
 \begin{align}
 \tilde{x}&=z \sin(k y)+x \cos(k y),\\
 \tilde{y}&= y,\\
 \tilde{z}&=z \cos(k y)-x \sin(k y),
 \end{align}
 for the kicks about the $y$-axis with kicking strength $k$, and, with azimuthal angle $\phi$ (see main text)
 \begin{equation}
 \tilde{\theta}=\arccos\left( \frac{1-(\frac{1-z}{1+z} \exp(2\tau)) }{ 1+(\frac{1-z}{1+z})  \exp(2 \tau)} \right),
 \end{equation}
 
 \begin{equation}
 \tilde{x}=\sin(\tilde{\theta}) \cos(\phi),
 \end{equation}
 \begin{equation}
 \tilde{y}= \sin(\tilde{\theta}) \sin(\phi),
 \end{equation}
 \begin{equation}
 \tilde{z}=\cos(\tilde{\theta}),
 \end{equation}
 for the superradiant damping, where
 \begin{equation}
 \tau = (2j+1) \gamma_\text{sr} t,
 \end{equation}
 for a time $t$, spin size $j$, and superradiant decoherence rate $\gamma_\text{sr}$.

\section{A closer look at the kicks set by the reinforcement learning agent}\label{app:distribution_of_kicks}

Here we take a closer look at the kicks chosen by the RL agent in the examples with superradiant damping, considered in Fig.~\ref{fig:superradiant_strategies} in the main text.

\begin{figure}[h!]
	\centering
	(a)
	\includegraphics[width=1\linewidth]{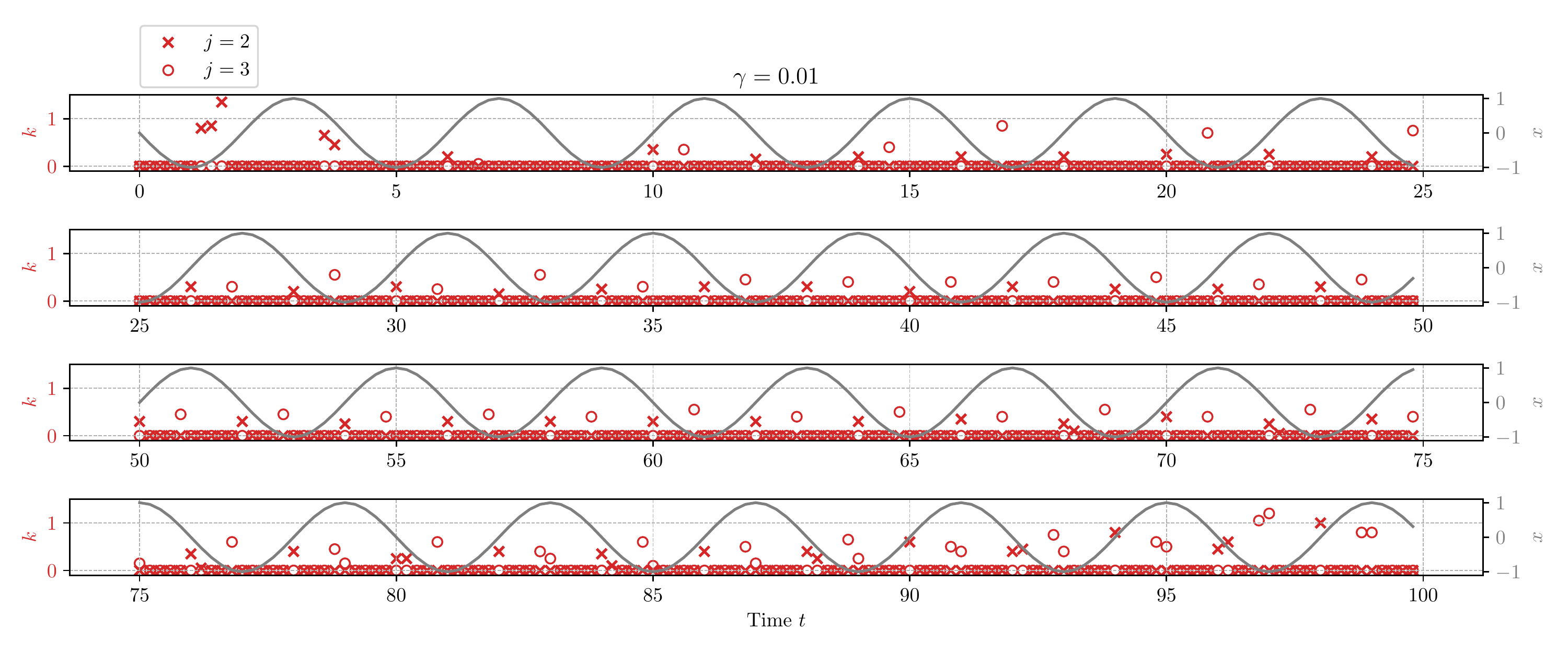}
	
	(b)
	\includegraphics[width=1\linewidth]{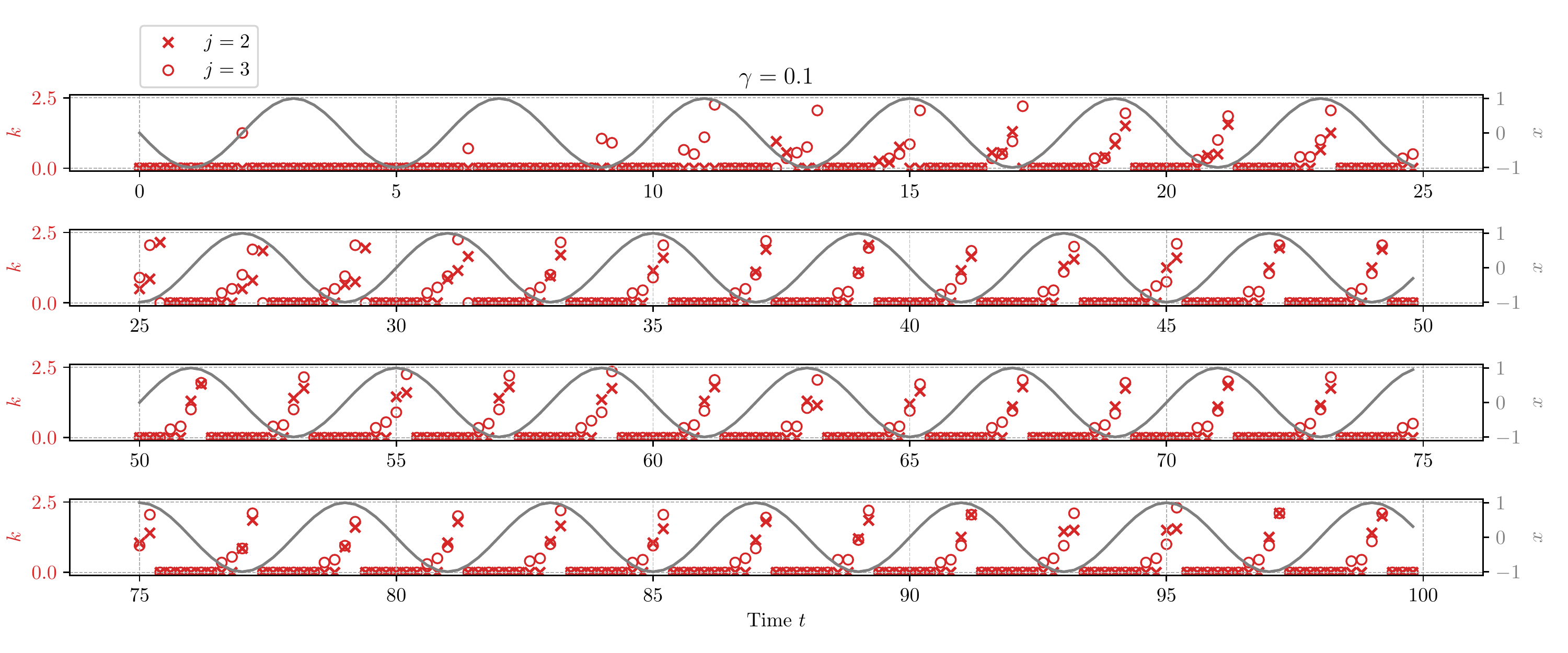}
	\caption{Kicks set by the RL agent for the SR-GKT. Panel (a) shows the case for $\gamma_\text{sr}=0.01$ and panel (b) for $\gamma_\text{sr}=0.1$. In red on the left axis are the kicking strengths \lu{$k$} for $j=2$ (crosses) and $j=3$ (circles). To show the precession, we plot on the right axis in grey the $x$ component of an unkicked spin coherent state without decoherence.}
	\label{fig:kicks}
	
\end{figure}

\begin{figure}[h!]
	\centering
	\includegraphics[width=0.8\linewidth]{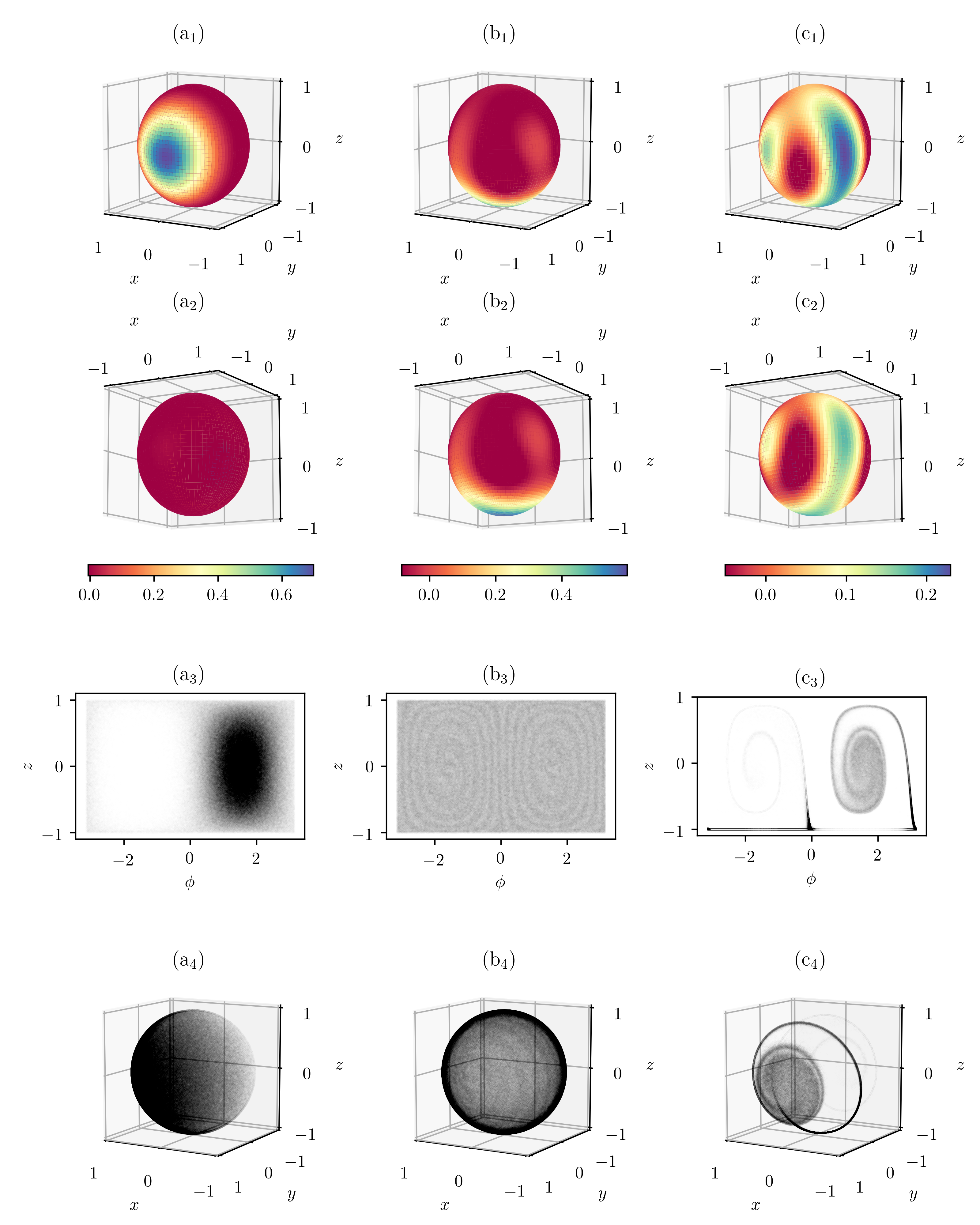}
	\caption{Shows the corresponding data as in Fig.~\ref{fig:classical_and_wigner} but for $j=2$ instead of $j=3$: Illustration of kicked superradiant dynamics with Wigner functions and its classical limit. The dissipation rate is $\gamma_\text{sr}=0.01$. Panels in the left column (a) correspond to the initial spin coherent state at $\theta=\phi=\pi/2$. The middle and right columns correspond to the state at time $T_\text{opt}$ generated with periodic kicks [middle column (b), $k=30$] and with kicks optimized with reinforcement learning [right column (c), the corresponding QFI is shown in panel (b) of Fig.~\ref{fig:superradiant_strategies}].  The top two rows show the Wigner functions of the density matrix, the bottom two rows show the classical phase space, populated by $10^6$ points initially distributed according to the Husimi distribution of the initial spin coherent state and then propagated according to the classical equations of motion.}
	\label{fig:classical_and_wigner2}
\end{figure}

In case of $\gamma_\text{sr}=0.01$, for both, $j=2$ and $j=3$, we find relatively similar distribution of kicks, see panel (a) in Fig.~\ref{fig:kicks}.
The most striking difference between the two policies for $j=2$ and $j=3$ are the comparatively strong kicks in the beginning of the sequence. By observing the time evolution of the Wigner function (see Supplemental Material), we find that these kicks basically rotate the state by an additional angle $\pi/2$ about the $z$-axis. This leads to a phase shift of $\pi/2$ between the two policies [see panels (d$_3$) and (d$_4$) of Fig.~\ref{fig:classical_and_wigner2}] compared to the initial state [see panels (a$_3$) and (a$_4$) of Fig.~\ref{fig:classical_and_wigner2}]. 

For $\gamma_\text{sr}=0.1$ the policies are even more similar with several kicks increasing in strength with a period length of $\pi$, see panel (b) in Fig.~\ref{fig:kicks}.

Fig. \ref{fig:classical_and_wigner2} is analog to Fig. \ref{fig:classical_and_wigner} in the main text but for $j=2$ instead of $j=3$.
The only qualitative difference compared to the $j=3$, is the periodically kicked top:
The combination of periodic kicks with $k=30$ and $j=2$ seems to be a special configuration. The classical phase space is comparable with the $j=3$ case, but there is much less structure in the Wigner function. Instead, the state concentrates on the south pole and exhibits a slightly squeezed shape (this is difficult to judge from Fig. \ref{fig:classical_and_wigner2} though). The rather high value of the QFI for $k=30$ and $j=2$, is best explained by this squeezing. When choosing other kicking strength, we observed a Wigner function similar to the case of $j=3$. 

\clearpage

\end{document}